# Electron modulation and ultrafast near-field imaging with vectorial laser fields

J. Kuttruff[†,*], L. Möhrle[†], L. Ciorciaro, L. Schmidt-Mende, P. Baum*

*Universität Konstanz, Fachbereich Physik, 78464 Konstanz, Germany*

[†]*equal contribution*

**joel.kuttruff@uni-konstanz.de*

**peter.baum@uni-konstanz.de*

Controlled interaction of laser light with electron beams is fundamental for ultrafast electron microscopy and electron-based quantum optics, yet their direct coupling is forbidden in free space. Here we use longitudinally polarized light at a thin membrane and show that the emerging focal fields can modulate the electron beam in a direct, coherent and linear way, without the need for nanostructured materials or slanted interaction geometries. Also, we use vectorial polarizations to excite and probe three-dimensional nanophotonic near-fields in metallic mesocrystals by coherent electron energy gain and loss. We find that longitudinal electric fields excite axial near-fields in a direct way while longitudinal magnetic fields excite oscillating ring currents via azimuthal electric fields. These possibilities enable tilt-free, collinear generation of attosecond electron pulses or free-electron qubits and provide novel imaging modes in ultrafast electron microscopy and metamaterial tomography.

Electrons and photons are two fundamental elementary particles and both are available in free space as bright beams. Understanding and controlling their interaction provides insight into basic quantum physics [1-4] but also enables advances in ultrafast electron microscopy [5,6], particle accelerators [7], or quantum optics [8-10]. However, the direct interaction of a laser photon with a free electron is very weak, as momentum and energy are not simultaneously conserved. Investigations therefore either need two photons per electron [11-13] or a rigid third body that is usually a nanostructure [2,3,8] or a thin membrane [14]. In addition, the electric field of the laser beam must have a longitudinal component in direction of the electron beam, in order to generate coherent energy gain or loss [2,3] that in turn modulates the electron pulses in the time domain [15]. Likewise, when aiming for attosecond imaging of nanophotonic materials [6,16,17] or determination of local intensities [3,18-20], we need a longitudinal excitation of the structure under scrutiny, so far achieved only indirectly by near-field conversion effects.

Here, we investigate how electrons interact with longitudinally or azimuthally polarized laser light and show that they can obtain energy gain and loss in a direct way. Figure 1a shows the concept of our experiments. We use an ultrafast transmission electron microscope with a Schottky field emitter (JEM F200, JEOL) at an electron energy of 200 keV and an electron pulse duration of ~270 fs [21]. The laser pulses (Carbide, LightConversion) have 1030-nm central wavelength, 250-fs pulse duration, and 2-MHz repetition rate. Part of the laser beam is frequency doubled and emits femtosecond electrons [22] with less than one electron per pulse [23]. Another part of the laser beam is used for electron modulation. We apply a liquid-crystal polarizer (ARCoptix) for reorientation of the electric field direction via a twisted nematic effect and half-wave plates [24]. This arrangement can switch the polarization state of the laser beam from linear (Fig. 1b) to azimuthal (Fig. 1c) or radial (Fig. 1d). An off-axis parabolic mirror (focal distance $f$ = 2.65 mm, numerical aperture NA = 0.35) with a 150-μm-radius hole in the middle focuses the shaped laser light collinearly with the electron beam onto a 50-nm-thick SiN membrane at the sample position of the electron microscope. The optical power is 35 μW and the focus diameter (full width at half maximum) is ~2 μm [25], only about two times the wavelength. The membrane serves as a passive interaction element that couples longitudinal electric fields to forward momentum of the electrons [26]. An electron spectrometer (CEFID, CEOS) and fast camera (Timepix, ASI) then record electron energy spectra or energy-filtered images on basis of single detected electrons [27]. Figure 1b shows the scenario without liquid crystal polarizer (linear polarization). In the center of the optical focus, the electric field (red arrows) and magnetic field (green arrows) are perpendicular to the k-vector, preventing energy transfer to the electrons (blue). Figure 1c shows the scenario for an azimuthal polarization of the incoming laser beam (faint red arrows). At the focus, we now generate a longitudinal magnetic field (green arrow) while the electric field vectors (red) point tangentially around the center of the beam. Therefore, we again expect no energy change of the electrons. Figure 1c shows the scenario for a radial polarization of the incoming beam (faint red arrows) in which the electric field vectors point outwards with anti-aligned phases at opposing positions around the center. When this beam is focused, we generate an oscillating longitudinal electric field (red). In the center, magnetic fields are zero for all times. We therefore expect an efficient electron-photon coupling in time and energy at zero transverse deflection by magnetic effects, in contrast to a tilted linear-polarized geometry [14,26,28,29]. In all cases, the effective acceleration or deceleration of the electrons will be given by a path integration over the optical cycles of the incoming, reflected and transmitted laser waves [26].

In the experiment, we record energy-filtered transmission electron microscopy images (EFTEM) at a magnification of 750 of the focused laser beam to capture and analyse not only the central fields but the entire focal plane. Coherent light-electron interaction is expected to produce energy sidebands at integer multiples of the photon energy [2,3]. To measure real-space images of this energy gain or loss, we set the energy filter to select only those electrons that have gained at least one quantum of photon energy [3]. At low coupling strength, current in the first sideband is quadratically proportional to cycle-averaged electric field strength [8], and we therefore directly see the squared amplitude $|E_z(x,y)|^2$ of the effective, time-integrated longitudinal field that interacts with the electrons. For theoretical reference, we calculate the electric and magnetic fields $E_{x,y,z}(x,y)$ and $B_{x,y,z}(x,y)$ in the focus plane with the Richards–Wolf vectorial diffraction method [30,31]. Azimuthally and radially polarized laser beams have a phase singularity at the center, implying zero intensity. We therefore model the spatial distribution of the incident field with a cylindrical vector beam (CVB) [32], assuming a ~1-mm-sized Gaussian beam shape ($1/e^2$ diameter of the light intensity) for the linear polarization and a ~2-mm sized beam ($4\sigma$ intensity) for the azimuthal and radial vector beams.

Figure 2 summarizes our experimental and theoretical results. For the case of linear polarization along the *y*-axis, Fig. 2a shows the incident electric field intensity before focusing, and Fig. 2b shows the *z*-component of the squared electric field amplitude at the focus plane. As expected, the longitudinal electric field vanishes at the center of the beam due to destructive interference of the partial rays on either side of the incident beam. The two lobes along the y-axis come from unbalanced destructive interference in our high-NA focusing [32]. Figure 2c shows the measured EFTEM images of the electron pulses that have interacted with this light field. We observe the expected zero energy gain and loss in the middle while two large sidelobes feature the same distribution and symmetry as in the simulation results. Figure 2d shows the measured electron energy spectrum at the center of the light beam, obtained by focusing the incident electron beam to a diameter of ~200 nm and setting the energy analyzer to spectroscopy mode. Around the zero-loss peak, we observe almost no light-induced spectral broadening and can describe the measured spectrum (dots) with a single-lobed Voigt function (solid line). Remaining small broadening is due to slight overlap of the electron beam with the two side lobes (Fig. 2c).

Figure 2e shows the intensity and field directions of an incoming laser beam with azimuthal polarization. In such a beam, the time-frozen electric fields point around the beam in azimuthal direction (black arrows). One half-cycle period later, the absolute direction is reversed. Figure 2f shows the simulated longitudinal squared electric field amplitude $|E_z|^2$ at the focal plane. We see $|E_z(x,y)|^2 = 0$ everywhere, because the focusing mirror cannot tilt the electric polarization anywhere around the beam into a longitudinal field. Figure 2g shows the measured EFTEM image for this configuration. We observe no light-induced energy gain. A weak inhomogeneous background in the lower-left part is attributed to imperfect iso-chromaticity alignment in our energy filter. Figure 2h shows the measured electron energy spectrum in the center of the light field. Again, we observe no light-induced spectral broadening, and the data (dots) again fits well to a single-lobed Voigt function (solid line).

Figure 2i shows the incident beam profile and time-frozen field vectors for the case of radial polarization. The beam shape is the same as for the azimuthal case, but all polarization vectors are now rotated by 90° to point inwards-out (black arrows). The focusing mirror now most efficiently converts these local polarizations into longitudinal fields. Figure 2j shows the simulated *z*-component of the squared electric field amplitude in the focal plane. As expected, we see a sharp spot that is smaller than the diffraction limit [33]. Figure 2k shows the measured EFTEM image of electrons that have interacted with these longitudinal fields. We observe one central intensity lobe, matching in position and size to the theory results. Figure 2l shows the measured electron energy spectrum. In contrast to the two previous cases, we now observe strong spectral broadening and several photonic sidebands [3]. Due to quantum path interference of the free-electron wave function with different optical cycles, this spectrum is proportional to Bessel functions [3,8] and the solid line shows a corresponding fit. We achieve a coupling strength [3,8] of $g \approx 0.7$ at a pulse energy of merely 17 pJ.

The agreement of all measured data with the theoretical amplitude of the longitudinal electric field in the focus plane shows that electron energy gain and loss are directly related to these fields, and not to membrane-induced projections of other fields into longitudinal direction, like in angled geometries [26]. The match also shows that the more complicated electromagnetic fields before and after the optical focus cancel out. Field oscillations average out in free space [2], but the membrane material introduces an optical phase shift of ~0.3 rad in our experiment that breaks the symmetry between incoming and transmitted waves, leaving predominantly a contribution of the fields close-by and inside the membrane to be observed [14,26]. This mechanism explains why we see in the experiments only and directly the longitudinal fields of the focal plane.

Magnetic fields cannot change kinetic energy but may deflect the beam. In none of the three experiments, we find any sideways deflection of the electron beam. In the case of linear polarization, this reveals the theoretically predicted cancelation between electric and magnetic contributions at velocity matching angles at normal incidence, that is, 90° between the electron beam direction $v_e$ and the membrane [26]. By contrast, the azimuthal and radial cases have, in the center, $B \parallel v_e$ and $B = 0$, respectively, cancelling sideways deflection effects in a direct way. The electron beam therefore always remains perpendicular to the crystallographic axis of a specimen during the interaction with the membrane and attosecond diffraction will be free of rocking-curve effects [34].From the measured quantum interference (Fig. 2l) and appearance of discrete photon-order sidebands that match similar results obtained with slanted excitation fields at nanostructures [3,8], we conclude that the light-electron coupling in our longitudinal electric fields is fully coherent and linear, as much as allowed by the finite temporal coherence length of the electrons [35]. The emerging attosecond electron pulses [11,36,37] or free-electron qubits [38] will therefore have no pulse front tilt [39].

Tomographic reconstruction of three-dimensional optical fields with attosecond electron microscopy [6] and related interferometry techniques [17,16] requires angled excitation of a nanostructure with arbitrarily polarized electromagnetic fields [40]. This is now possible with the reported scheme. Figure 3 shows energy-filtered electron microscopy data of the interaction of our polarization-shaped fields with different assemblies of metallic nanostructures. We drop a suspension of gold nanoparticles of cubic shape with ~60 nm size and organic ligands [41] onto a 50-nm SiN membrane and let them self-assemble into mesocrystals, that is, small clusters with preferential order between adjacent particles [42]. Figure 3a shows a transmission electron microscopy image with bright-field contrast of one such cluster, consisting of roughly ten cubes. Lighter areas correspond to a single layer of particles while darker areas indicate stacking regions where several cubic particles attach to each other along the *z*-axis. The shadow-like features are remaining ligands at the sides of the individual particles. In a first experiment (Fig. 3b), we excite this nano-assembly with linearly polarized light (Fig. 1b) with an electric polarization in the *x*-*y*-plane at 45° (see inset) and look for *z*-components with EFTEM. We observe two main lobes which appear at 45° at a size that is ~10 times smaller than the residual lobes of the incoming field (compare Fig. 2c). We argue that the dipolar modes of each individual nanoparticle produce a collective, enhanced excitation of the entire cluster, producing one mesoscopic dipole mode. Figure 3c shows the results for an azimuthal focal field with |E| = 0 and B || z (see inset). Although direct electron-photon coupling is zero in this case (Fig. 2g-h), we observe non-zero intensity in EFTEM imaging, particularly at sharp edges or protruding tips of the clustered nanoparticle assembly. Electric fields are zero in the middle but rotate azimuthally around the center (see inset). We argue that these fields, even if weak at nanometer distances, drive a circular current in our mesoscopic cluster that in turn localizes and converts to longitudinal electric fields at certain edges of the non-round particle. These longitudinal components then couple to the electrons, resulting in measurable but weak energy exchange. Figure 3d shows the results for radial polarization and longitudinal electric fields (see inset). A largely homogeneous intensity around the nanoparticle cluster comes from the SiN membrane (compare Fig. 2k-l). On the top side of the nanostructure ($y > 0$ nm), we observe an enhancement of this modulation, and on the bottom side ($y < 0$ nm), we observe a quenching of the longitudinal fields, producing regions with no coupling at all (dashed black lines). We argue that the interaction of individual nanoparticle modes, now polarized along the $z$-axis and highly sensitive to thickness and number of cubes, produces constructive or destructive optical interferences, or longitudinal surfing effects [6] that amplify or hinder energy transfer to the electrons. Such longitudinal effects are not observed for linear excitation and therefore a direct manifestation of the vectorial character of the focal field.

Figures 3e-h show a similar experiment at a larger and more complex mesocrystalline material with a fractal-like geometry. Figure 3e shows a bright-field image of the area of interest, showing several islands and half-islands with combinations of single-layered and stacked regions. Figure 3f shows the nanophotonic resonances for a linearly polarized excitation field at 45° in the $x$-$y$-plane (see inset). We observe regions with higher intensity, again mostly oriented along the polarization direction. However, in some regions, the shape of the mesostructure breaks the symmetry and we observe intensity lobes oriented in a tilted way; see for example top-right pear-like shape at $x \approx 300$ nm and $y \approx 300$ nm. Figure 3g shows the results for azimuthal polarization and no electric field at the center (compare Fig. 1c). Residual circular fields induce ring currents that in turn localize by near-field effects to longitudinal electric fields, visible as high-intensity regions in proximity of larger sub-clusters (e.g. at $x \approx -200$ nm, $y \approx 400$ nm) or in semi-closed voids (e.g. at $x \approx 300$ nm, $y \approx 0$ nm) where circularly oscillating fields can resonantly propagate. Several such resonances can constructively or destructively interfere, causing regions of zero intensity between two nano-islands (see dashed line). Figure 3h shows the results for the radially polarized case with longitudinal electric fields. Again, we see a homogeneous background from the interaction with the flat membrane, but it is now modulated by regions of enhanced intensity and regions of zero

intensity (dashed lines), caused by longitudinal surfing effects [6] or optical interference of different longitudinal near-field modes that oscillate forth and back along the *z*-axis.

The demonstrated capability for excitation of nanostructures with longitudinal electric fields will be useful, for example, to investigate vertically aligned nanostructures [32], tip-enhanced Raman sensing [43] or enhanced second-harmonic generation in nanocones [44] or nanowires [45]. Azimuthal focal fields can reveal higher-order multipole contributions to nanoparticle scattering that are otherwise hidden under standard illumination [46]. The strong longitudinal component of the magnetic field in the focus of an azimuthally polarized beam can be used to selectively drive magnetic modes at optical frequencies [47] or magnetic Mie resonances [48,49]. Such magnetic fields will also enable novel tests of the ultrafast Aharonov-Bohm effect [50]. In laser-electron control, the reported longitudinal electric fields in radially polarized beams can lead to optimized photon-electron coupling and allow for direct, efficient and tilt-free modulation of an electron beam into attosecond pulses or free-electron qubits. Inserting the reported longitudinal fields into a TM-like waveguide mode, for example in a photonic crystal or glass fiber, will match the phase velocity of the light to the speed of the electrons over an extended range, potentially providing an electron-photon exchange strength approaching unity [51], a key requirement for utilizing coherent electron-photon interactions for free-electron quantum optics or quantum electron microscopy.

**Acknowledgements:** We thank Stefan Schupp for help with mesocrystal preparation. This work was supported by the Deutsche Forschungsgemeinschaft (project 510996696 and SFB 1432) and by the European Research Council (AdG ULMI).

**Author contributions:** All authors performed research and wrote the manuscript.

**Data availability:** The data supporting the findings of this study are available from the corresponding author upon request.

**Disclosures:** The authors declare no conflicts of interest.

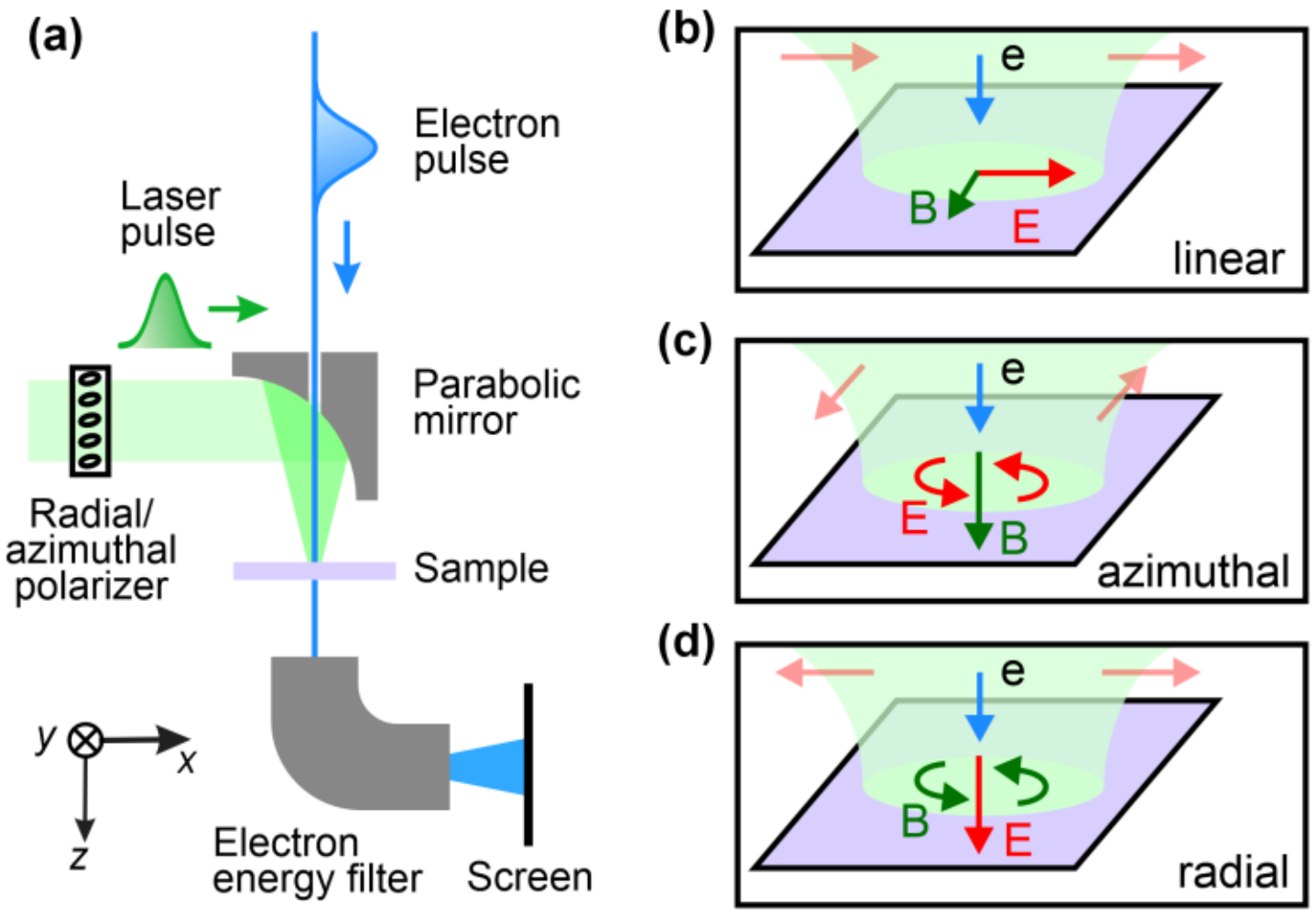

**Fig. 1. Setup and vectorial focal fields.** (a) In an ultrafast electron microscope, a laser beam (green) is polarization-shaped by a liquid-crystal radial/azimuthal polarizer (rectangle with ellipses) and then tightly focused on a sample (purple). The electron pulses (blue) are overlapped with the laser beam and become modulated by the electromagnetic fields. An energy filter records the electron spectrum and produces energy-filtered images of the laser-driven specimen. (b) Illustration of the electromagnetic fields for a linear incoming polarization (faint red arrows). In the focus plane (purple), this linear polarization is mostly maintained (red and green arrows). (c) Electromagnetic field for an azimuthal incoming polarization (faint red arrows). The electric field in the middle becomes zero and oscillates around the center ray (red arrows). The magnetic field becomes longitudinal (green arrow). (d) Electromagnetic fields for radial polarization (faint red arrows). In the focus, we produce a longitudinal electric field (red), surrounded by a circular magnetic field (green arrows). For multi-cycle pulses, all arrows quickly change sign at the frequency of the laser wave.

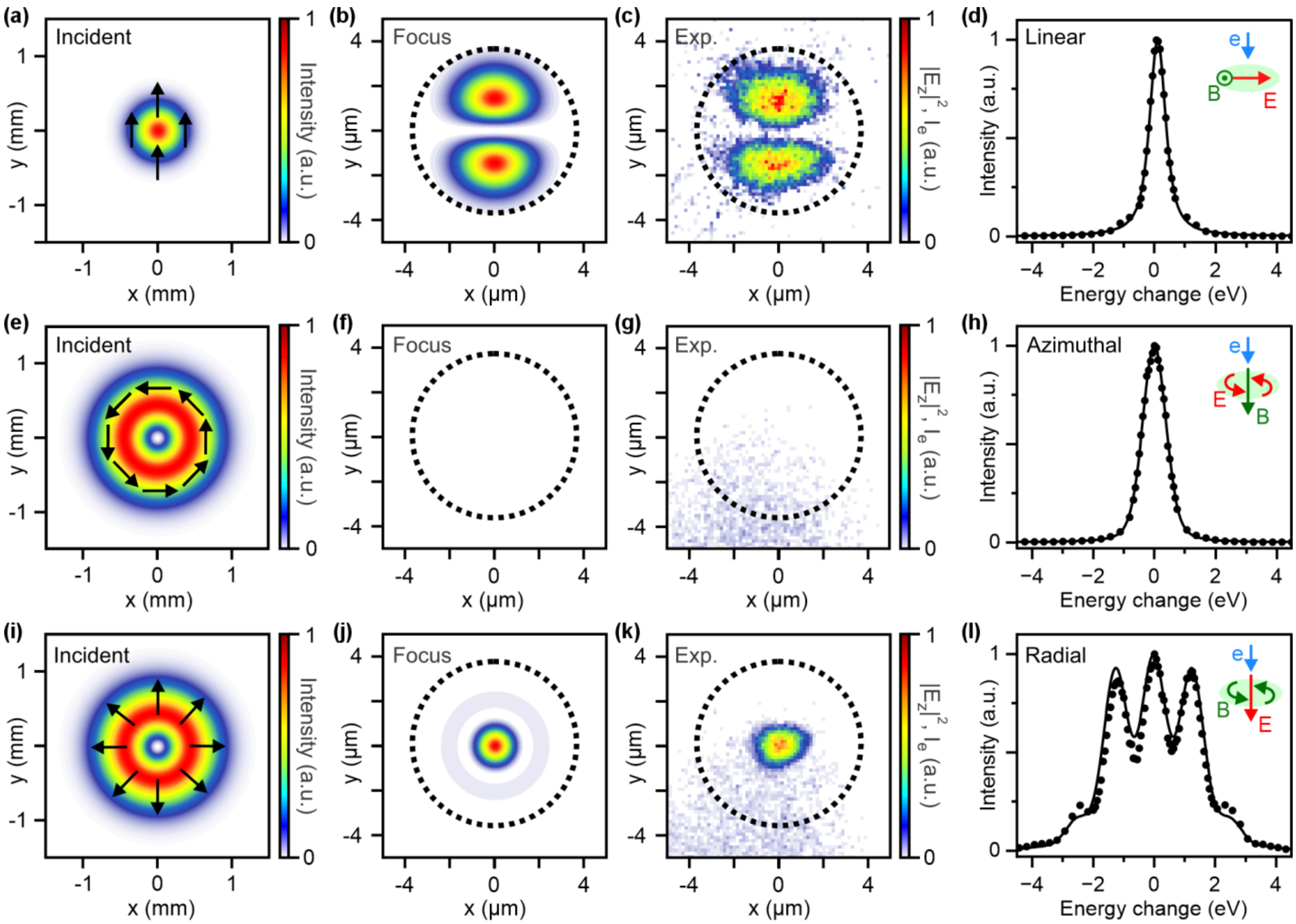

**Fig. 2. Electron modulation with vectorial focal fields.** (a) Simulated spatial profile of the incident laser beam (colored data) for linear polarization along the *y*-axis (black arrows). (b) Focal distribution of the longitudinal electric field, cycle-averaged and squared. The two intensity lobes are oriented along the *y*-axis. There is no longitudinal component in the center of the beam. Dashed circle, guide to the eye. (c) Measured energy-filtered electron image (EFTEM) after interaction with the linear-polarized focal field. (d) Measured electron energy spectrum (dots). Inset, field direction; solid line, Voigt fit. (e) Simulated incident beam profile for azimuthal incoming polarization (black arrows). (f) Focal distribution of the longitudinal electric field, cycle-averaged and squared. The field is zero everywhere. Dashed circle, guide to the eye. (g) Measured EFTEM image after interaction with the azimuthal focal field. Remaining intensity is due to imperfect iso-chromaticity of the experiment. (h) Measured electron energy spectrum (dots) after interaction with an azimuthal focal field (inset). Solid line, Voigt fit. (i) Simulated incident beam profile for radial incoming polarization (black arrows). (j) Focal distribution of the longitudinal electric field, cycle-averaged and squared. The spot size is below the diffraction limit for a Gaussian beam. Dashed circle, guide to the eye. (k) Measured EFTEM image after interaction with a radial focal field. (l) Measured electron energy spectrum (dots) after

interaction with a radial focal field (inset). Electrons are coherently accelerated and decelerated in the longitudinal electric fields. Solid line, coherent interaction model (see text).

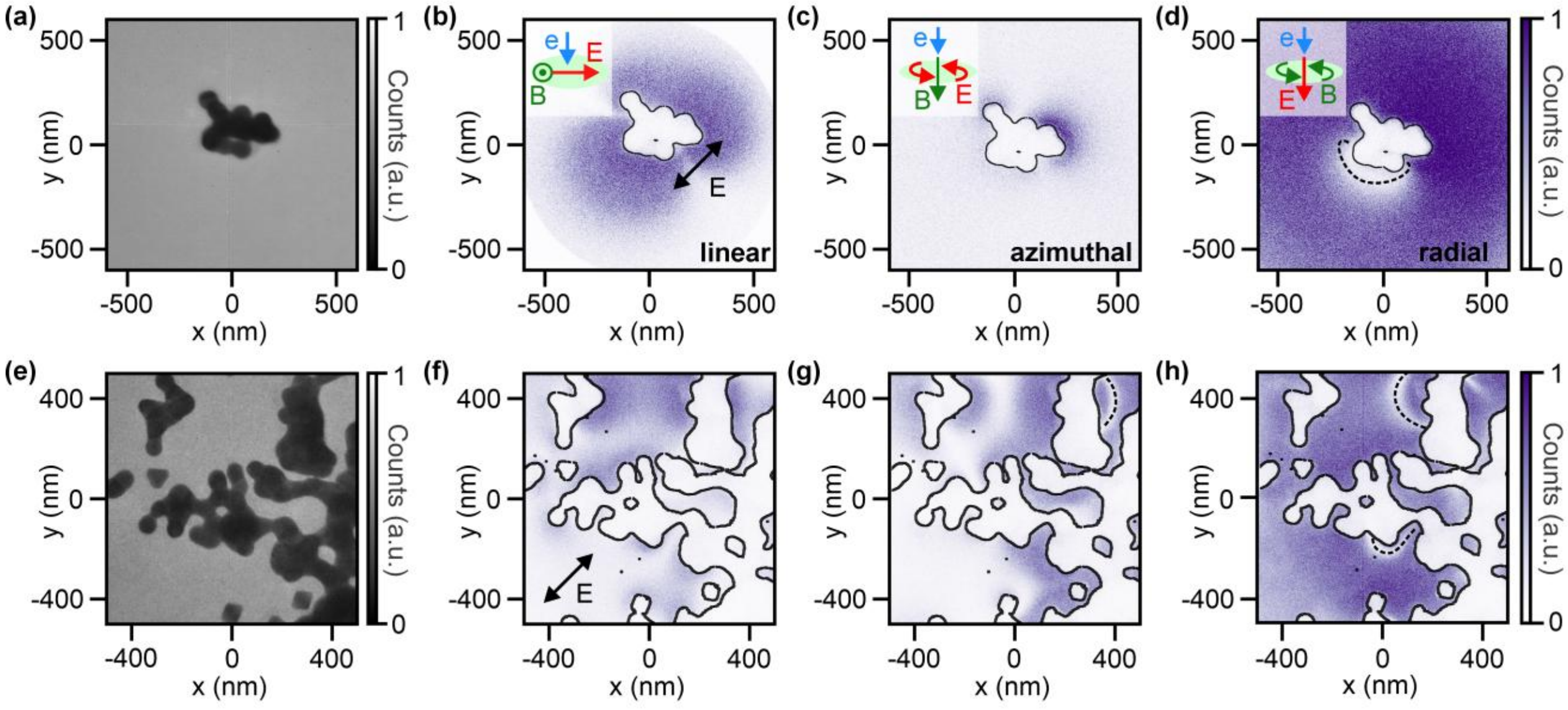

**Fig. 3. Interaction of vectorial focal fields with metallic mesostructures.** (a) Transmission electron microscopy image of a mesocrystal cluster, consisting of several self-assembled gold nanoparticles. Darker areas indicate stacking of nanoparticles. (b) Energy-filtered electron image (EFTEM) after interaction with linearly polarized light (black arrow). The dashed line indicates the outline of the cluster. (c) EFTEM image after interaction of the mesocrystal with an azimuthal focal field. (d) EFTEM image after interaction of the mesocrystal with a radial focal field. The dashed line indicates regions with zero intensity from longitudinal interference effects. (e) Transmission electron microscopy image of a more complex mesocrystal assembly. (f-h) EFTEM images after interaction with a linear, azimuthal and radial focal field, respectively. Arrows and dashed lines denote polarizations and areas of destructive longitudinal interference, respectively.